\documentclass[preprint]{aastex}
\usepackage[]{natbib}
\usepackage[]{verbatim}
\usepackage[]{float}

\newcommand{\feh}{\ensuremath{\rm [Fe/H]}}

\newcommand{\teff}{T$_{\rm eff}$}
\newcommand{\logl}{log(L/L$_{\odot}$)}

\newcommand{\kms}{km\ s$^{-1}$}
\newcommand{\mv}{M$_{\rm V}$}

\newcommand{\msun}{M$_{\odot}$}

\newcommand{\vsini}{$v$\,sin\,$i$}
\newcommand{\loggf}{log($gf$)}
\newcommand{\logg}{log($g$)}
\newcommand{\vmt}{$v_{\rm mt}$}
\newcommand{\vk}{($V-K_{s}$)}
\newcommand{\loglxlbol}{log(L$_{\rm x}$/L$_{\rm bol})$}
\newcommand{\lxlbol}{log(L$_{x}$/L$_{bol})$}

\slugcomment{submitted to Astrophysical Journal}
\shorttitle{AB Dor Group}
\shortauthors{Barenfeld et al.}
\begin{document}

\title{A Kine-Chemical Investigation of the AB Dor Moving Group
  ``Stream''}

\author{Scott A. Barenfeld\altaffilmark{1}, Eric J.
  Bubar\altaffilmark{1,2}, Eric E. Mamajek\altaffilmark{1,3}, Patrick
  A. Young\altaffilmark{4}}
\altaffiltext{1}{University of Rochester, Department of Physics \& Astronomy,
  Rochester, NY, 14627-0171, USA}
\altaffiltext{2}{Marymount University, Department of Biology and Physical Sciences, Arlington, VA 22207-4299,
USA}
\altaffiltext{3}{Cerro Tololo Inter-American Observatory, Casilla 603, La Serena, Chile}
\altaffiltext{4}{Arizona State University, School of Earth and Space Exploration,
Tempe, AZ 85287-1404, USA}

\begin{abstract}
  The AB Dor Moving Group consists of a ``nucleus'' of $\sim$10 stars
  at $d$ $\simeq$ 20 pc, along with dozens of purported ``stream''
  members distributed across the sky.  We perform a chemical and
  kinematic analysis of a subsample of AB Dor stream stars to test
  whether they constitute a physical stellar group. We use the NEMO
  Galactic kinematic code to investigate the orbits of the stream
  members, and perform a chemical abundance analysis using high
  resolution spectra taken with the Magellan Clay 6.5-m telescope.
  Using a $\chi^2$ test with the measured abundances for 10 different
  elements, we find that only half of the purported AB Dor stream
  members could possibly constitute a statistically chemically
  homogeneous sample. Some stream members with 3D velocities were
  hundreds of parsecs from the AB Dor nucleus $\sim$10$^8$ yr ago, and
  hence were unlikely to share a common origin.  We conclude that the
  published lists of AB Dor moving group stream members are unlikely
  to represent the dispersed remnant of a single star formation
  episode. A subsample of the stream stars appears to be both
  statistically chemically homogeneous and in the vicinity of the AB
  Dor nucleus at birth.  Their mean metallicity is [Fe/H] =
  0.02\,$\pm$\,0.02 dex, which we consider representative for the AB
  Dor group.  Finally, we report a strong lower limit on the age of
  the AB Dor nucleus of $>$110 Myr based on the pre-MS contraction
  times for K-type members which have reached the main sequence.
\end{abstract}


\section{Introduction}
\label{section:intro}

It has long been recognized that the solar neighborhood contains a
population of young stars with ages and velocities similar to the
Pleiades, including the famous star AB Dor
\citep[e.g.][]{Jeffries1995}.  More recently, \citet{Zuckerman2004}
identified a concentration of stars associated with AB Dor at $d$
$\simeq$ 20 pc of apparently similar age and velocity.  Using galactic
3D space velocities and youth indicators such as H$\alpha$ emission,
strong Li absorption, strong X-ray emission, fast rotation, and
color-magnitude diagram position, they identified 37 candidate member
systems of the ``AB Dor Moving Group'', 9 of which appear to comprise
a ``nucleus,'' including AB Dor itself. A detailed examination of the
age indicators for AB Dor and its group members by \citet{Luhman2005}
convincingly demonstrated that the color-magnitude diagram and Li
depletion pattern for the group is suggestive of coevality with the
Pleiades open cluster (hence a probable age of $\sim$125 Myr). In the
years since, additional members of the AB Dor moving group have been
proposed \citep{Torres2008, VianaAlmeida2009, Schlieder2010,
  Zuckerman2011}.  \citet{DaSilva2009} have tested the membership of
these stars, and proposed new members, using an iterative method involving
the proximity of candidate stars to each other in UVWXYZ space and to
an adopted isochrone in absolute visual magnitude \citep[see
also][]{Torres2006}.

If the stars in a moving group are to have shared a common origin,
then like open clusters, they should exhibit not only similar space
velocity, but also similar chemical composition.  \citet{DeSilva2007a} and
\citet{Bubar2010} examined the HR 1614 and Wolf 630 moving groups,
respectively, and have shown that using kinematics alone to group
stars can be unreliable.  These studies found that stars previously
identified as group members based solely on their kinematics did not
match the abundance patterns exhibited by the other stars in the
group.  Moving groups such as these with larger velocity spread
(usually called ``{\it superclusters}'') are now believed to be
created by dynamical perturbations caused by, e.g., the Galactic bar.
They clearly have a wide ranges of ages \citep{Famaey2008, Bovy2010},
and hence are {\it not} useful samples for age-related studies of
stars.  When adopting ages for stars based on their membership to a
kinematic group, it is important to know whether the group is
consistent with being co-eval and co-chemical.

Considering the utility of moving groups for understanding galactic
kinematic and chemical evolution, and their interest as targets for
planet imaging and circumstellar disk evolution surveys, we have
started a project to ``chemically tag'' some of the young, nearby
stellar groups. In this contribution, we test whether purported
members of the AB Dor moving group could have a shared origin. We
present a detailed kinematic and spectroscopic study of 10 stars
identified by \citet{Torres2008} and \citet{DaSilva2009} as AB Dor
members. These authors have already demonstrated that these stars have
Li abundances consistent with other AB Dor members, so we do not
discuss Li further. The 10 stars are a subsample of the ``stream''
members with low projected rotational velocity (\vsini\, $<$ 20 \kms)
that are outside of the ``nucleus'' identified by
\citet{Zuckerman2004}.

\section{Observations and Reduction \label{sec:obs}}

High resolution optical echelle spectra of the AB Dor stream stars
listed in Table \ref{tab:pp} were obtained on the nights of June
25-26, 2010 with the MIKE spectrograph at the 6.5m Clay telescope at
Las Campanas Observatory.  Data was reduced using standard procedures
in the IRAF echelle package.  These include bias correction, flat
fielding, scattered light removal and wavelength calibration.  The
tilted slits were dealt with using the IRAF {\sf mtools} package.  In
order to assure measurement of clean, unblended spectral features, we
limited our abundance analysis to the red CCD, therefore the resultant
spectra have wavelength coverage from 4850-8500 \AA,
with a resolution of R $\simeq$ 60,000 and typical S/N $\simeq$
200-300 per res. element.  For reference, our analysis was carried out
with respect to an extremely high S/N solar spectrum from reflected
light from the asteroid Ceres, measured with the same telescope and
setup.

\section{Analysis \label{sec:anal}}

\subsection{Spectroscopic Analysis}
\label{section:spec-analysis}

We followed a standard excitation/ionization balance approach to
determine basic physical parameters from our stellar spectra
\citep{Bubar2010}. For our initial guesses of these parameters, we
used photometric temperatures using the calibrations of
\citet{Casagrande2010} and gravities from the tracks of
\citet{Baraffe1998}.  Our input metallicity was assumed to be solar,
and we calculated \vmt\, using equation 2 from
\citet{AllendePrieto2004}.

The largest sources of error in our abundances were uncertainties in
the final physical parameters and uncertainties in the line
measurements themselves.  Other sources of error, such as the \loggf\,
values used in the line lists are eliminated (to first order) by our
use of a differential abundance analysis.  Uncertainties in {\feh} and
the other physical parameters of each star were found using the method
described in \citet{Bubar2010}.  Because this method gave us
unrealistically large uncertainties in \logg, we adopted the \logg\,
uncertainties using the \citet{Baraffe1998} evolutionary tracks.
The physical parameters are given in Table \ref{tab:pp}.  In addition,
we include measurements of the equivalent widths of the lithium doublet at
6707\AA.  Our measurements agree with those from \cite{DaSilva2009}.

We also measured lines of \ion{Na}{1}, \ion{Mg}{1}, \ion{Al}{1},
\ion{Si}{1}, \ion{Ca}{1}, \ion{Cr}{1}, \ion{Mn}{1}, \ion{Ni}{1}, and
\ion{Ba}{2}.  Using the temperature, gravity, microturbulence, and
metallicity solution for each star, we determined the abundances
relative to the Sun, line by line.  We used the mean abundance as our
final value for each element and the standard error of the different
line abundances as our uncertainty.  When a single line was available,
we adopted a conservative uncertainty of 0.10 dex.  If more than one
line was available, but all lines gave the same value (standard error
of 0), we adopted a conservative uncertainty of 0.05 dex. The
abundances are listed in Table \ref{tab:abun}.


\begin{deluxetable}{llcclccccc}
\rotate
\tablecaption{Spectroscopic and Photometric Parameters of AB Dor Stream Stars\label{tab:pp}}
\tablewidth{0pt}
\tablehead{ 
\multicolumn{1}{c}{} & \multicolumn{3}{c}{Photometric} & \multicolumn{4}{c}{Spectroscopic} & \multicolumn{2}{c}{Activity}\\
 & T$_{\rm eff}$ & \logg & $v_{mt}$ & T$_{\rm eff}$ & \logg & $v_{mt}$ & EW(Li)& \lxlbol & $\Delta$EW(H$\alpha$)\\
Star & (K)           &(dex)  & (\kms)   & (K)           & (dex) & (\kms) & (m\AA) & (dex) & (\AA)}
\startdata
BD -03 4778    & 5045$\pm$102 & 4.47$\pm$0.08 & 1.11$\pm$0.04 & 5220$\pm$65 & 4.31$\pm$0.08 & 1.80$\pm$0.11 & 261 & -3.38 & -4.46\\
HD 6569        & 5080$\pm$87 & 4.60$\pm$0.07 & 1.10$\pm$0.03 & 5170$\pm$59 & 4.61$\pm$0.07 & 1.37$\pm$0.12  & 141 & -3.79 & -3.36\\
HD 189285      & 5685$\pm$119 & 4.32$\pm$0.07 & 1.40$\pm$0.06 & 5537$\pm$56 & 4.46$\pm$0.07 & 1.51$\pm$0.09 & 136 & -3.86 & -0.58\\
HD 199058      & 5647$\pm$93 & 4.20$\pm$0.05 & 1.43$\pm$0.05 & 5737$\pm$71 & 4.62$\pm$0.05 & 1.05$\pm$0.12  & 152 & -3.93 & -0.48\\
HD 207278      & 5615$\pm$101 & 4.45$\pm$0.06 & 1.32$\pm$0.04 & 5710$\pm$60 & 4.56$\pm$0.06 & 1.70$\pm$0.10 & 188 & -3.96 & -0.63\\
HD 217343      & 5761$\pm$99 & 4.43$\pm$0.06 & 1.38$\pm$0.04 & 5830$\pm$59 & 4.59$\pm$0.06 & 1.70$\pm$0.10  & 165 & -4.12 & -0.90\\
HD 218860A     & 5488$\pm$91 & 4.54$\pm$0.07 & 1.24$\pm$0.04 & 5543$\pm$49 & 4.59$\pm$0.07 & 1.45$\pm$0.08  & 216 & -3.61 & 0.40\\
HD 224228      & 4876$\pm$79 & 4.63$\pm$0.06 & 1.03$\pm$0.02 & 4953$\pm$52 & 4.56$\pm$0.06 & 1.11$\pm$0.13  &  70 & -4.36 & -3.02\\
HD 317617      & 4570$\pm$142 & 4.50$\pm$0.10 & 0.94$\pm$0.05 & 4870$\pm$63 & 4.49$\pm$0.10 & 1.10$\pm$0.15 & 109 & -3.92 & 0.19\\
TYC 486-4943-1 & 4680$\pm$271 & \nodata       & \nodata      & 5160$\pm$81 & 4.87$\pm$0.10 & 2.50$\pm$0.21  & 178 & -3.12 & -6.64
\enddata
\tablecomments{EW(Li) is the equivalent width of the \ion{Li}{1} doublet feature at 6707\AA. $\Delta$EW(H$\alpha$) is the residual chromospheric H$\alpha$ emission measured by subtracting a template spectrum of an inactive star of similar \teff\, and approximately solar composition (see \S4).}
\end{deluxetable}


\begin{deluxetable}{lrrrrrrrrrr}
\rotate
\tablecaption{Abundances of AB Dor Stream Stars\label{tab:abun}}
\tablewidth{0pt}
\tablehead{Star & [Na/H] & [Mg/H] & [Al/H]  & [Si/H]  & [Ca/H]  & [Cr/H]  & [Mn/H]  & [Ni/H]  & [Ba/H] & [Fe/H] }
\startdata
BD -03 4778    & -0.03(5) &  -0.16(6) & -0.05(4) & -0.15(3) &  0.06(4) & 0.10(6) & -0.14(2) & -0.18(2) &  0.03(3) &  -0.09(4)\\
HD 6569        & -0.05(5) &  -0.02(6) &  0.01(2) & -0.04(3) &  0.04(2) & 0.20(2) &  0.02(3) &  0.00(3) &  0.22(2) &   0.06(3)\\
HD 189285      & -0.12(5) &  -0.11(7) & -0.05(3) & -0.07(3) &  0.03(3) & 0.00(10)& -0.12(2) & -0.12(3) &  0.11(1) &  -0.03(4)\\
HD 199058      & -0.08(3) &  -0.15(12)& -0.14(1) & -0.06(3) &  0.01(4) & 0.00(10)& -0.09(4) & -0.13(2) &  0.19(2) &  -0.03(5)\\
HD 207278      &  0.01(2) &  -0.14(10)&  0.01(3) &  0.02(4) &  0.09(3) & 0.18(15)& -0.07(1) & -0.06(2) &  0.25(3) &   0.02(5)\\
HD 217343      & -0.08(4) &  -0.15(4) &  0.08(10) & -0.04(3) & -0.01(3) & 0.17(18)& -0.18(4) & -0.11(3) &  0.18(3) &  -0.04(4)\\
HD 218860A     & -0.06(4) &   0.00(8) &  0.02(6)  &  0.02(2) &  0.09(4) & 0.14(5) & -0.01(3) & -0.02(2) &  0.26(2) &   0.05(3)\\
HD 224228      & -0.08(8) &  -0.12(3) & -0.14(10) & -0.09(3) &  0.07(8) & 0.07(1) & -0.04(1) & -0.09(2) &  0.12(2) &  -0.04(3)\\
HD 317617      &  0.01(9) &  -0.04(7) & -0.04(4)   & -0.10(3) &  0.11(9) & 0.10(5) & -0.09(1) & -0.08(1) &  0.10(5) &  -0.03(3)\\
TYC 486-4943-1 &  0.05(10) & \nodata &  0.07(2) & -0.39(9) &  0.04(8) & 0.06(7) & -0.14(1) & -0.12(5)&  0.00(5) &  -0.10(5)\\
\enddata
\tablecomments{Values in parentheses are 1$\sigma$ uncertainties in
  final digits. A recent study by \citet{D'Orazi12} measured [BaII/Fe]
  = 0.10$\pm$0.15 dex for HD 189285 (TYC 5155-1500-1), and [BaII/Fe] =
  0.20$\pm$0.15 dex and HD 218860 (HIP 114530), consistent with the
  supersolar [Ba/H] values that we measured.}
\end{deluxetable}

\subsection{Kinematics}
\label{section:kinematics}

\begin{figure}
\begin{center}
\includegraphics[scale=0.35,angle=-90]{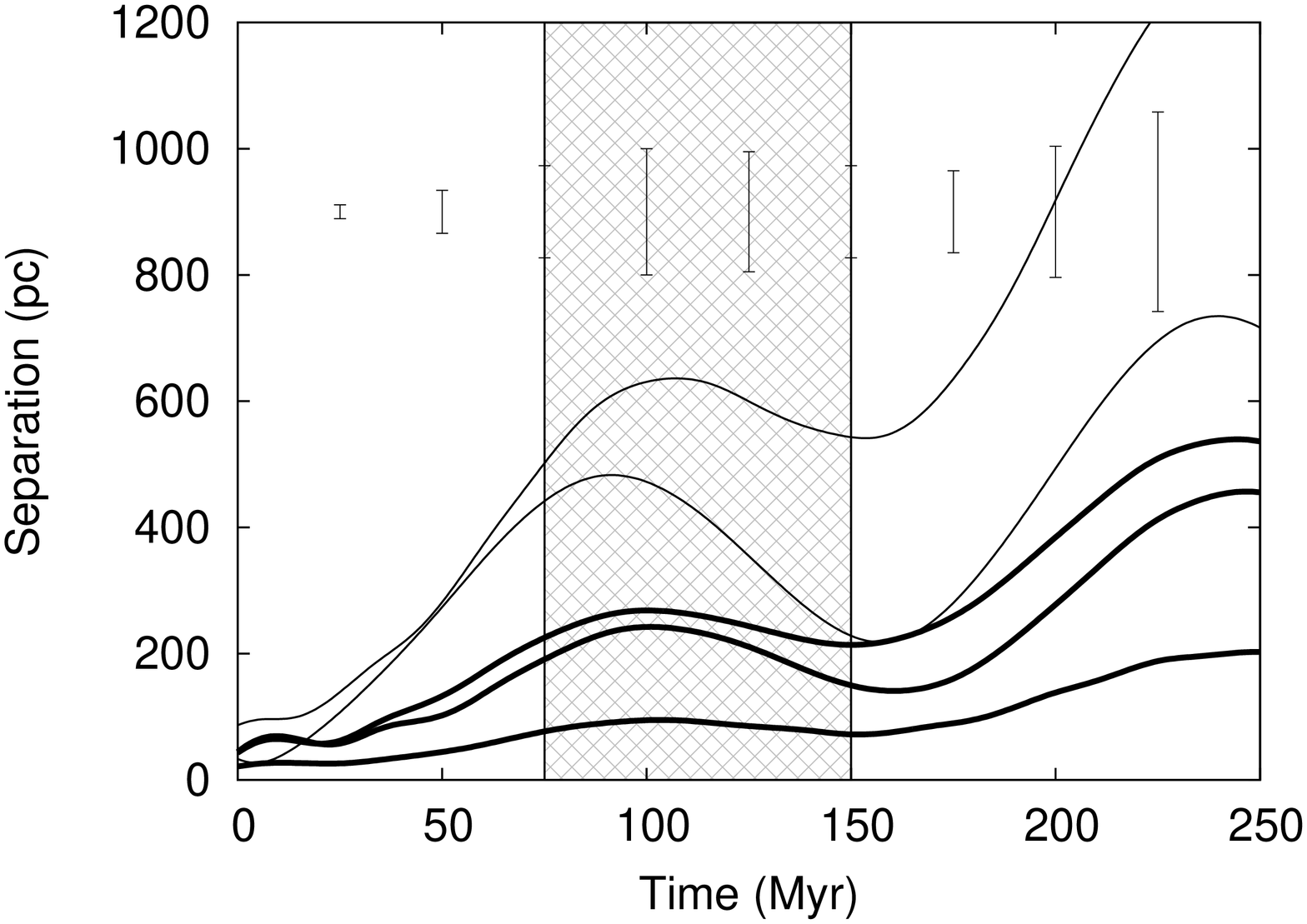}
\caption{Separations between the 5 AB Dor stream members with
  velocities listed in Table \ref{tab:uvw} and the AB Dor nucleus over
  the past 250 Myr. The {\it hatched region} corresponds to the plausible 
  age range for the AB Dor nucleus (70-150 Myr, most likely near
  $\sim$125 Myr; but see discussion in \S\ref{age}). 
  The three stars with thicker lines tracing their past separations
are discussed in \S4.  They appear to be statistically chemically
 homogeneous with one another, and could have formed together with
the AB Dor nucleus. Mean 1$\sigma$ uncertainties in separation 
are plotted at 25 Myr intervals at top.  The stars plotted here are, from top 
to bottom, HD 207278, HD 217343, HD 218860A, HD 6569, and HD 224228.
\label{fig:sep}
}
\end{center}
\end{figure}

For 5 of the 10 stars we studied spectroscopically, we calculated 3D
velocities using published astrometry and radial velocities and the
matrices of \citet{Johnson1987}. The other 5 stars are lacking
Hipparcos trigonometric parallaxes \citep{vanLeeuwen2007}. The
velocities\footnote{We follow the usual convention where U is the
  velocity towards the galactic center, V is the velocity in the
  direction of galactic rotation, and W is the velocity in the
  z-direction out of the galactic plane.} and references are
summarized in Table \ref{tab:uvw}.  We also include the mean velocity
of the AB Dor nucleus stars as calculated by \citet{Mamajek2010} and
Mamajek (in prep.) which combines astrometry from
\citet{vanLeeuwen2007} with the radial velocities of
\citet{Gontcharov2006} for the AB Dor nucleus members from
\citet{Zuckerman2004}.  \citet{Mamajek2010} and Mamajek (in prep.)
estimate the mean velocity of the AB Dor nucleus to be (U, V, W) =
(-7.6\,$\pm$\,0.4, 27.3\,$\pm$\,1.1, -14.9\,$\pm$0.3) with a 1D
velocity dispersion of 1.0\,$\pm$\,0.4 \kms.  We then used the
software package NEMO \citep{Teuben1995} and the Galactic potential
(model 2) of \citet{Dehnen1998} to calculate the past orbits of our
stream stars, and their separations from the nucleus as a function of
time over the past 250 Myr (roughly twice the likely age).  Though our
stars are currently within tens of pc of the nucleus, they disperse in
the past, and two are $>$400 pc away from the AB Dor nucleus for any
reasonable published estimate of its age (see Fig.  \ref{fig:sep}).
If the AB Dor group is indeed coeval with the Pleiades ($\sim$125
Myr), then 2 of our 5 stream stars with 3D velocities were
$\sim$400-600 pc away from the AB Dor nucleus when it was born (where
the separation uncertainties are $\sim$190 pc 125 Myr ago).  The
density of the AB Dor nucleus is too low\footnote{The AB Dor
  nucleus contains roughly $\sim$8 \msun\, of stars within a
  volume of $\sim$2500 pc$^{3}$, for a density of $\sim$0.003 \msun\,
  pc$^{-3}$ \citep{Mamajek2010}, which is a factor of $\sim$40 lower
  than the local disk density \citep[$\rho_0$ $\simeq$ 0.12 \msun\,
  pc$^{-3}$;][]{vanLeeuwen2007}. It is unclear whether this subset of
  members constitutes a true ``nucleus'' or whether it is a chance
  over-density of stream members.} to have influenced the orbits of
the stream members during this time.  {\it At these past separations,
  at least two of the five stream stars are unlikely to have formed in
  the same molecular cloud as the AB Dor nucleus}.


\begin{deluxetable}{l c c c c c c}
\tablecaption{Velocities of AB Dor Stream Stars\label{tab:uvw}}
\tablewidth{0pt}
\tablehead{ 
Star & U    & V    & W    & Ref.\\
...  & \kms & \kms & \kms & ...}
\startdata
HD 6569    & -7.9$\pm$1.2 & -28.9$\pm$1.2 & -10.0$\pm$1.2 & 1,4\\
HD 207278  & -8.3$\pm$3.0 & -30.2$\pm$2.5 & -12.8$\pm$1.6 & 2,4 \\
HD 217343  & -3.2$\pm$0.4 & -25.4$\pm$0.4 & -13.6$\pm$0.3 & 1,5\\
HD 218860A & -8.3$\pm$1.2 & -28.3$\pm$0.9 & -10.3$\pm$0.7 & 3,5\\
HD 224228  & -7.5$\pm$0.4 & -27.7$\pm$0.3 & -13.5$\pm$0.3 & 1,4\\
AB Dor nuc.& -7.6$\pm$0.4 & -27.3$\pm$1.1 & -14.9$\pm$0.3 & 6\\
Pleiades   & -6.8$\pm$0.6 & -28.1$\pm$0.6 & -14.1$\pm$0.4 & 7\\
\enddata
\tablecomments{All velocities calculated using parallaxes from
  \citet{vanLeeuwen2007}. Proper motion and radial velocity
  references: (1) \citet{vanLeeuwen2007}, (2) UCAC3
  \citep{Zacharias2010}, (3) PPMX \citep{Roeser2008}, (4)
  \citet{Torres2006}, (5) \citet{Nordstrom2004}, (6)
  \citet{Mamajek2010} and Mamajek (in prep.). (7) Velocity for the
  Pleiades was calculated using the mean cluster proper motion from
  \citet{vanLeeuwen2007}, parallax from \citet{Soderblom2005}, and
  radial velocity from \citet{Robichon1999}.}
\end{deluxetable}

\subsection{An Age Constraint on AB Dor \label{age}}

\begin{figure}[H]
\begin{center}
\includegraphics[scale=0.3]{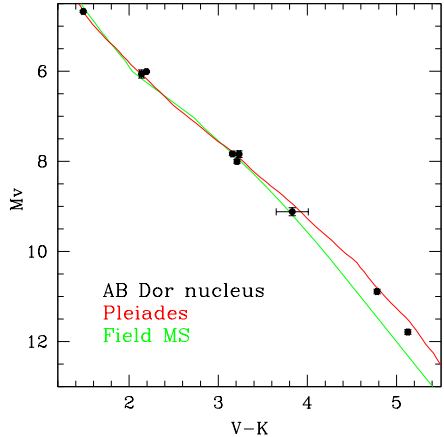}
\caption{Color-magnitude diagram for members of the
AB Dor nucleus, using photometry on the Johnson V and 2MASS K$_s$ 
systems. The three late K-type stars
HIP 25283, 26369, and 31878 are clustered near
(V-K$_s$, \mv $\simeq$ 3.2, 7.9). Note that the
Pleiades empirical isochrone strongly overlaps the
field star main sequence for V-K$_s$ $<$ 3.4. Using the
\citet{Baraffe1998} tracks, this point corresponds
roughly to a pre-main sequence contraction age of
$\sim$120 Myr (for a 0.65 \msun\, star).
\label{fig:colormag}
}
\end{center}
\end{figure}

One can constrain the age of the AB Dor nucleus by searching for the
main sequence turn-on point \citep[e.g.][]{Pecaut2012}. Using the
\citet{Zuckerman2004} sample of nucleus stars, we constructed a
V-K$_s$ vs. M$_V$ color-magnitude diagram (Fig. \ref{fig:colormag})
using V magnitudes from \citet{ESA1997}, K$_s$ photometry from 2MASS
\citep{Skrutskie2006}, and parallaxes from \citet{vanLeeuwen2007}. We
compare the positions of the AB Dor nucleus stars to the dereddened
Pleiades color-magnitude sequence from \citet{Stauffer2007}, and a
main sequence constructed using the relations of \citet{Wright2005}
and a custom B-V vs. V-K$_s$ color-color fit for field
stars\footnote{http://www.pas.rochester.edu/$\sim$emamajek/EEM\_dwarf\_UBVIJHK\_colors\_Teff.dat}.

The Pleiades and field star main sequence are in remarkable agreement
for stars blueward of V-K$_s$ $<$ 3.4 \citep[\teff\, $\simeq$ 4000 K;
using calibration of ][]{Casagrande2008} and M$_V$ $<$ 8.3 (\logl\,
$>$ -1.02). Using the \citet{Baraffe1998} tracks, this turn-on ZAMS
position corresponds to a 0.65 \msun\, star. It takes such a star 120
Myr to contract as a pre-MS star before reaching within 0.01 dex
luminosity of the ZAMS. This agrees remarkably well with other modern
turnoff and Li depletion ages for the Pleiades \citep[$\sim$125 Myr;
e.g.][]{Stauffer1998, Barrado2004, Kharchenko2005}.

For the AB Dor nucleus, there is a well-defined clump of late K-type
members which appear to be on the ZAMS: including HIP 25283 (\vk,
\mv\, = 3.16, 7.84), HIP 26369 (\vk, \mv\, = 3.23, 7.92) \& HIP 31878
(\vk, \mv\, = 3.21, 8.00). The color magnitude diagram is sparse
redward of this, however the known nucleus members redward of V-K$_s$
$>$ 4.78 (\teff\, $<$ 3180 K; HIP 22738A, B, AB Dor Ba/Bb) are
certainly pre-MS. {\it Considering its color-magnitude position with
  respect to other nucleus members, AB Dor itself is clearly ZAMS, not
  pre-MS, as commonly quoted} (\vk, \mv\, = 2.26, 6.05)\footnote{Based
  on the multidecadal V-band photometry presented in Fig. 2 
  of \citet{Innis2008}, we adopt an average V magnitude of AB Dor of
  6.95.}. Hence the use of pre-MS evolutionary tracks for AB Dor A is
inappropriate. The stars blueward of V-K$_s$ $<$ 3.25 (\teff\, $>$
4140 K) and brighter than M$_V$ $<$ 8.0 (\logl\, $>$ -0.96) are
definitely on the MS. Using the \citet{Baraffe1998} tracks, this
corresponds to stars of mass $>$0.67 \msun. It takes a 0.67 \msun\,
star 110 Myr to contract to the ZAMS, {\it hence we can take 110 Myr
  as a strong lower limit on the age of the AB Dor nucleus}.

The cooler AB Dor nucleus members HIP 22738 A (V-K$_s$ = 4.78, M$_V$ =
10.89) and HIP 22738 B (V-K$_s$ = 5.13, M$_V$ = 11.79) have
color-magnitude positions nearly coincident with the single-star
Pleiades sequence of \citet{Stauffer2007}, when one adopts the HST
Pleiades distance from \citet{Soderblom2005}. The location of ZAMS
stars with V-K$_s$ $\simeq$ 3.2 is at odds with previous younger age
estimates \citep[50-70 Myr;][]{Zuckerman2004, Torres2008,
  DaSilva2009}, and with recent age estimates for the AB Dor system
itself of 40-50 Myr \citep[][]{Guirado2011}\footnote{However these
  authors conceded that ``Older ages are not completely excluded by
  our data''} and 50-100 Myr \citep{Janson2007}. Our results further
corroborate the findings of \citet{Luhman2005} that the AB Dor group
is coeval with the Pleiades ($\sim$125 Myr).


\section{Discussion}

A stellar group that formed simultaneously within a molecular cloud is
expected to be chemically homogeneous, except for elements potentially
depleted as the stars age (e.g. Li, Be).  To test for chemical
homogeneity within our sample, we developed an abundance $\chi^2$
test.  We calculated a $\chi^2$ value for each star individually,
using $\chi^2_{star} = \Sigma\frac{(X_i-\hat{X}_i)^2}{\sigma_i^2}$
where $X_i$ is the measured abundance of the $i$th element, $\sigma_i$
is the uncertainty in this measurement, and $\hat{X}_i$ is the
expected abundance of the star, obtained by a weighted linear least
squares fit to the abundance vs. \teff\, trend for each element.  The
$\chi^2$ values of the individual stars were then summed to obtain a
total $\chi^2$, $\chi^2_{tot}$.  We compared $\chi^2_{tot}$ to the
95\%-significance critical values.  If our stars constitute a
chemically homogeneous sample, $\chi^2_{tot}$ should be less than the
critical value.  If $\chi^2_{tot}$ was too high, we rejected the star
with the highest individual $\chi^2$, and repeated the above
calculations with the remaining stars.  We continued this iterative
procedure until a statistically homogeneous sub-sample was found.  Out
of our original sample of 10 stars, we found only 5 to be consistent
with being chemically homogeneous: HD 189285, 224228, 217343, 199058
and 317617 (Table \ref{tab:chi}). {\it Our results suggest that
  roughly half of the purported AB Dor stream stars have dissimilar
  chemical compositions}.


\begin{deluxetable}{llrrrrrrrrrrr}
\rotate
\tablecaption{A Chemically Homogeneous Subsample of AB Dor Stream Stars\label{tab:chi}}
\tablewidth{0pt}
\tablehead{HD&\teff\ (K) &[Na/H]&[Mg/H]&[Al/H] &[Si/H] &[Ca/H] &[Cr/H] &[Mn/H] &[Ni/H] &[Ba/H]&[Fe/H]&$\chi^2$}
\startdata
317617&4870(63) & 0.01(9)&-0.04(7)&-0.04(4)&-0.10(3) &0.11(9) &0.10(5)&-0.09(1)&-0.08(1) &0.10(5)&-0.03(3) &8.6\\
224228&4953(52) &-0.08(8)&-0.12(3)&-0.14(10)&-0.09(3) &0.07(8) &0.07(1)&-0.04(1)&-0.09(2) &0.12(2)&-0.04(3) & 11.3\\
217343&5830(59) &-0.08(4)&-0.15(4)& 0.08(10)&-0.04(3) &-0.01(3) &0.17(18)&-0.18(4)&-0.11(3)&0.18(3)&-0.04(4) &8.9\\
199058&5737(71) &-0.08(3)&-0.15(12)&-0.14(1)&-0.06(3) &0.01(4) &0.00(10)&-0.09(4)&-0.13(2)&0.19(2)&-0.03(5) &7.2\\
189285&5537(56) &-0.12(5)&-0.11(7)&-0.05(3)&-0.07(3) &0.03(3) &0.00(10)&-0.12(2)&-0.12(3) &0.11(1)&-0.03(4)&9.7\\
\enddata
\tablecomments{Values in parentheses are 1$\sigma$ uncertainties in final digits.}
\end{deluxetable}

To quantify the degree of chemical heterogeneity of the stream sample,
we also calculated the intrinsic abundance scatter necessary to
generate the observed scatter for each element in our sample.  We
mirrored the approach of \citet{DeSilva2006}, in that
$\sigma_{obs}^2=\sigma_{int}^2+\sigma_{err}^2$, where $\sigma_{obs}$
is our observed standard deviation from the mean abundance of each
element, $\sigma_{err}$ is the average measurement uncertainty we have
for the abundance of each element, and $\sigma_{int}$ is the intrinsic
scatter we are solving for.  We find intrinsic 1$\sigma$ dispersions
of 0.02 (Na, Mg), 0.03 (Fe), 0.04 (Cr, Ni), 0.06 (Al, Mn), 0.08 (Ba),
and 0.11 (Si). The observed scatter in the Ca abundances is consistent
with no intrinsic dispersion for the sample. \citet{DeSilva2006} found
that the intrinsic dispersion (rms) in abundances for Hyades cluster
members was typically $<$0.03 dex, and most notably only 0.014 dex for
Ba (compared to 0.08 dex seen for AB Dor stream stars) and scatter
consistent with no intrinsic dispersion for Si (compared to 0.11 dex
seen for AB Dor stream stars). This scatter is also consistent with
the HR diagram for the Hyades, which \citet{Quillen02} found yields an
intrinsic scatter in [Fe/H] of $<$0.03 dex rms.  {\it The intrinsic
  scatter in abundances for AB Dor stream stars is larger than that
  for a typical open cluster like the Hyades}.

\begin{figure}[H]
\begin{center}
\includegraphics[scale=0.35]{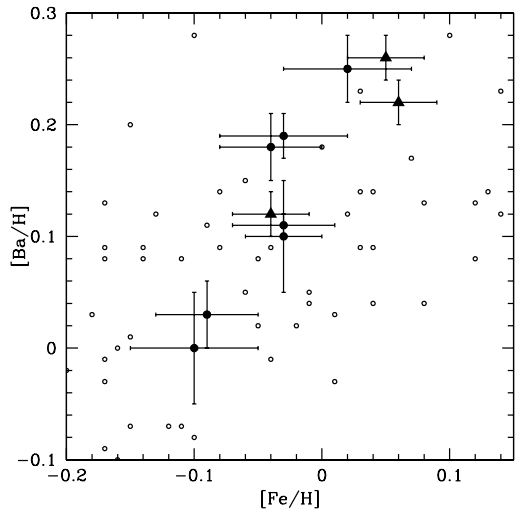}
\caption{Comparison of [Fe/H] vs. [Ba/H] abundances
for AB Dor stream stars ({\it large filled circles}) and 
field stars within 15 pc of the Sun from
\citet{AllendePrieto2004} ({\it small open circles}).  The three stars 
plotted with triangles are chemically and kinematically coherent (see \S4).
\label{fig:feh}
}
\end{center}
\end{figure}

We also compare our abundance results to abundances of field stars
within 15 pc of the Sun from the S$^4$N survey of
\citet{AllendePrieto2004}.  In Fig. \ref{fig:feh}, we plot [Fe/H] vs.
      [Ba/H] for our stars and the S$^4$N field stars.  Our stars
      qualitatively match the abundance trends of the field stars. The
      other elements our sample has in common with the S$^4$N survey
      show similar results.  Our results are suggestive that the AB
      Dor stream stars comprise a sample of young stars with a range
      of chemical compositions.

One possible explanation for the observed scatter in our abundances is
stellar activity \citep[e.g.][]{Schuler2010}.  All of the stars have
X-ray counterparts in the ROSAT All-Sky Survey \citep{Voges99}, so we
calculate X-ray fluxes following \citet{Fleming95} and quote the
coronal activity of the stars as \loglxlbol\, in Table \ref{tab:pp}.
We calculate Spearman correlation coefficients for coronal activity
(\loglxlbol) vs. abundances for the 10 elements we investigated among
the AB Dor stream stars (see Table \ref{tab:SpearLx}). None of the
trends are statistically significant (adopting $\alpha$ = 0.05
significance level), as determined using a critical value table
\citep{Zar72}.  We also test whether abundance trends exist vs.
chromospheric activity as quantified using H$\alpha$ emission.  We
calculate residual H$\alpha$ equivalent widths by subtracting
normalized spectra of similar resolution of stars of similar
temperature and approximately solar metallicity from the
\citet{Montes98} library of echelle spectra.  These residuals are
listed in Table \ref{tab:pp}.  In Table \ref{tab:SpearHa}, we list
the Spearman rank order correlation cofficients for residual H$\alpha$
emission equivalent widths versus elemental abundances.  [Mg/H] and
[Fe/H] show $\sim$2$\sigma$ correlations, while our other abundances
show no significant correlations with H$\alpha$ residuals.


\begin{deluxetable}{lr}
\tablecaption{Spearman Rank Order Correlation Coefficients for \loglxlbol\, vs. Abundance 
\label{tab:SpearLx}}
\tablewidth{0pt}
\tablehead{\loglxlbol  & $\rho$\\
             vs.       & } 
\startdata
  $[$Na/H$]$  &  0.48\\
  $[$Mg/H$]$  & -0.10\\
  $[$Al/H$]$  &  0.28\\
  $[$Si/H$]$  & -0.30\\
  $[$Ca/H$]$  &  0.12\\
  $[$Cr/H$]$  & -0.12\\
  $[$Mn/H$]$  & -0.10\\
  $[$Ni/H$]$  & -0.18\\
  $[$Ba/H$]$  & -0.33\\
  $[$Fe/H$]$  & -0.06
\enddata
\tablecomments{For two-tailed test and sample of 10 objects, the
  $\alpha$ = 0.05 level of significance corresponds to $\rho$ =
  $\pm$0.648 \citep{Zar72}.}
\end{deluxetable}


\begin{deluxetable}{cccc}
\tablecaption{Spearman Rank Order Correlation Coefficients for $\Delta$EW(H$\alpha$) vs. Abundance 
\label{tab:SpearHa}}
\tablewidth{0pt}
\tablehead{
\multicolumn{2}{c}{Full Sample} & \multicolumn{2}{c}{TYC 486-4943-1 and BD-03 4778 removed}\\
$\Delta$EW(H$\alpha$) vs. & $\rho$ & $\Delta$EW(H$\alpha$) vs. & $\rho$}
\startdata
  $[$Na/H$]$  & -0.30 & $[$Na/H$]$  &  0.05\\
  $[$Mg/H$]$  &  0.62 & $[$Mg/H$]$  &  0.26\\
  $[$Al/H$]$  & -0.12 & $[$Al/H$]$  & -0.07\\
  $[$Si/H$]$  &  0.45 & $[$Si/H$]$  &  0.07\\
  $[$Ca/H$]$  &  0.27 & $[$Ca/H$]$  &  0.43\\
  $[$Cr/H$]$  & -0.04 & $[$Cr/H$]$  & -0.40\\
  $[$Mn/H$]$  &  0.30 & $[$Mn/H$]$  & -0.19\\
  $[$Ni/H$]$  &  0.25 & $[$Ni/H$]$  & -0.14\\
  $[$Ba/H$]$  &  0.50 & $[$Ba/H$]$  &  0.02\\
  $[$Fe/H$]$  &  0.58 & $[$Fe/H$]$  &  0.12
\enddata
\tablecomments{$\Delta$EW(H$\alpha$) is the estimated chromospheric
  H$\alpha$ emission (see Table 1 \& \S4). For two-tailed test and
  sample of 10 objects, the $\alpha$ = 0.05 level of significance
  corresponds to $\rho$ = $\pm$0.648 \citep{Zar72}. Whether or not the
  two active stars TYC 486-4943-1 and BD-03 4778 are included in the
  sample, none of the activity vs. abundance correlations have
  significance beyond $\alpha$ = 0.05.}
\end{deluxetable}

However, before making conclusions regarding the intrinsic scatter in
the elemental abundances and trends of abundances vs. activity, we
note that two stars (BD-03 4778 and TYC 486-4943-1) are substantially
more active than the other stars in our sample (both in terms of
coronal X-ray emission and chromospheric H$\alpha$ emission; see Table
\ref{tab:pp}). When these stars are removed from calculation of the
Spearman correlation coefficients for H$\alpha$ emission
vs. abundance, we do not see any statistically significant
correlations. To examine the influence of these two stars on our
results, we repeat our calculation of the intrinsic elemental
abundance scatters without these two stars. We find intrinsic
1$\sigma$ dispersions of 0.01 dex (Fe, Mg), 0.02 dex (Si), 0.04 dex
(Ni), 0.05 dex (Cr), 0.06 dex (Mn, Ba), 0.07 dex (Al), and again
negligible scatter for the Ca and Na abundances. The intrinsic scatter
for [Si/H] dropped significantly when the two active stars are removed
from the sample.  While the scatter is negligible for some elements,
it is measurably higher for others (e.g. Cr, Mn, Ba, Al) than one
would expect for a cluster sample. {\it We thus conclude that activity
  alone is unlikely to explain the observed heterogeneity in
  abundances among the AB Dor stream stars.}


\begin{deluxetable}{llrrrrrrrrrrr}
\rotate
\tablecaption{A Chemically and Kinematically Coherent Population of AB Dor Stream Stars\label{tab:pop}}
\tablewidth{0pt}
\tablehead{HD&\teff\ (K) &[Na/H]&[Mg/H]&[Al/H] &[Si/H] &[Ca/H] &[Cr/H] &[Mn/H] &[Ni/H] &[Ba/H]&[Fe/H]&$\chi^2$}
\startdata
224228&4953(52) &-0.08(8)&-0.12(3)&-0.14(10)&-0.09(3)&0.07(8)&0.07(1)&-0.04(1)&-0.09(2)&0.12(2)&-0.04(3)&5.0\\
218860A&5543(49)&-0.06(4)& 0.00(8)& 0.02(6) & 0.02(2)&0.09(4)&0.14(5)&-0.01(3)&-0.02(2)&0.26(2)& 0.05(3)&7.5\\
6569&5170(59)   &-0.05(5)&-0.02(6)& 0.01(2) &-0.04(3)&0.04(2)&0.20(2)& 0.02(3)& 0.00(3)&0.22(2)& 0.06(3)&17.5\\
Wt. Mean& ............. &-0.06(3)&-0.09(3)& 0.01(2) &-0.02(1)&0.05(2)&0.10(1)&-0.03(1)&-0.04(1)&0.20(1)& 0.02(2)& ........\\ 
\enddata
\tablecomments{Values in parentheses are 1$\sigma$ uncertainties in final digits.}
\end{deluxetable}


Besides the five stars listed in Table \ref{tab:chi}, we found that
another subsample of 3 stars (HD 6569, HD 224228, and HD 218860A) is
also consistent with chemical homogeneity. Interestingly, these 3
stars also stay within 200 pc of each other in the past (well closer
than any other pairs of stars).  These 3 stars, displayed in Table
\ref{tab:pop} and shown in bold in Fig. \ref{fig:sep}, are also the 3
closest stars to the AB Dor nucleus in the past.  This combination of
chemical and kinematic homogeneity indicates that these 3 stars could
have formed together in the same birthsite, along with AB Dor. Their
weighted mean metallicity is [Fe/H] = 0.02\,$\pm$\,0.02 dex, and if
one places the 10 elements on equal footing, one derives a mean
metallicity of [M/H] = 0.01\,$\pm$\,0.02 dex. This is nearly identical
to a previous mean estimate for the AB Dor group by \citet{Ortega2007}
([Fe/H] = -0.02\,$\pm$\,0.02 dex), and the average quoted metallicity
[Fe/H] for the Pleiades \citep[+0.04\,$\pm$\,0.02
  dex;][]{Soderblom2009}.  Our combined chemical and kinematic results
suggest that these values are most representative of the true AB Dor
group.  {\it Thus, we find that a subsample of the AB Dor stream stars
  may constitute a kinematically and chemically coherent population,
  but that one should not assume that all stream stars have a common
  origin with one another or the AB Dor nucleus.}

The nature of the group of 5 chemically homogeneous stars\footnote{It
  is more correct to say that they are statistically consistent with
  constituting a chemically homogeneous subsample within the levels of
  our abundance uncertainties.} in Table \ref{tab:chi} is more
difficult to determine.  Only 2 of these stars have Hipparcos
parallaxes, so we could not run all 5 through our kinematic tests.  It
is possible that some or all of these stars could represent their own
stream, distinct from AB Dor, but we cannot say this definitively
without more precise kinematic data.

\section{Summary}

We have obtained high-resolution spectra of 10 purported AB Dor moving
group ``stream'' members.  Using measured abundances for 10 elements
(including Fe) we show that our sample of stream stars is
statistically inconsistent with being chemically homogeneous.  The
abundance trends of these stars are consistent with field star trends,
and our results suggest that perhaps half of the stream stars can be
considered statistically chemically homogeneous, whereas the other
half show slightly different chemical compositions which could reflect
birth in regions other than AB Dor's birthsite.  Due to the lack of
statistical correlations between stellar activity indicators
(\loglxlbol\, and H$\alpha$ emission) and the individual stellar
abundances, we surmise that stellar activity alone is unable to
explain the observed spread in abundances.  Kinematically, only 5 of
our stars have well determined 3D velocities, but we find that 2 of
these were $\sim$400-600 pc away from the AB Dor nucleus when it was
born, whereas 3 of them (which also appear to be statistically
chemically homogeneous) could have formed in AB Dor's vicinity at the
group's birth. This kinematically and chemically coherent group has
mean metallicity of [Fe/H] = 0.02\,$\pm$\,0.02 dex, which we think is
representative of the AB Dor group.  While there does appear to be an
AB Dor ``nucleus'' \citep{Zuckerman2004}, it appears that a
significant fraction of the outlying stream members found in
\citet{Torres2008} and \citet{DaSilva2009} do not constitute a
chemically homogeneous or kinematically coherent sample.

We also demonstrate that the AB Dor nucleus must be $>$110 Myr based
on the presence of 3 late K-type members which are clearly on the
zero-age main sequence. This disagrees with the often-cited ages for
the AB Dor group and AB Dor multiple system of $\sim$40-100 Myr.
Taking into account the findings of \citet{Luhman2005}, the data are
strongly in favor of coevality of the AB Dor nucleus with the Pleiades
($\sim$125 Myr).

Our survey shows that kinematics, color-magnitude positions, and
stellar youth indicators alone are not necessarily sufficient for
testing whether a kinematic group of stars actually shares a common
origin.  Chemical tagging of purported members of moving groups
provides an additional diagnostic for testing group membership, and
holds promise for piecing together the recent chemo-kinematic history
of star-formation in the solar vicinity.

\acknowledgements 

This research was supported by NSF grant AST-1008908, an REU
supplement, and funds from the School of Arts and Sciences at the
University of Rochester. We thank the Arizona Telescope Allocation
Committee for their allocation of Magellan time, and the staff of Las
Campanas Observatory for their technical support and hospitality.  We
thank the referee John Stauffer for several suggestions which improved
the manuscript, and Mark Pecaut for comments.

{\it Facilities:} \facility{Magellan:Clay (MIKE spectrograph)}, \facility{HIPPARCOS}, \facility{ROSAT}, \facility{CTIO:2MASS}

\end{document}